\documentclass{pasj01}
\Received{2024/02/02}
\Accepted{2024/04/04}

\usepackage{url}
\usepackage{multirow}

 \def\deg{^\circ} 
 \def\deg{^\circ}
 \def\/{\over}\def\kms{km s$^{-1}$}

 \def\be{\begin{equation}} 
 \def\ee{\end{equation}}
 \def\kms{km s$^{-1}$} \def\Kkms{K \kms }   
    
\def\({\left(} \def\){\right)} \def\[{\left[} \def\]{\right]}

 \def\Tex{T_{\rm ex}}  
\def\Tb{T_{\rm B}}

 \def\Ncoth{N_{\rm ^{13}CO}}
\def\Ycoth{Y_{\rm ^{13}CO}}
\def\htwo{H$_2$} 
 
\def\uNHtwo{\htwo\ cm$^{-2}$\ }
\def\vlsr{V_{\rm LSR}} 
\def\Tbcotw{T_{\rm B}({\rm ^{12}CO})} 
\def\Tbcoth{T_{\rm B}({\rm ^{13}CO})}

\def\Xgeneral{X_{\rm CO}}

\def\kms{km s$^{-1}$}
\def\Xg{\Xgeneral}

\begin{document}

\title{The CO-to-H$_2$ Conversion Factor in the Central Molecular Zone of the Milky Way using CO isotopologues}
\author{Mikito Kohno\altaffilmark{1,2}$^{*}$}%
\altaffiltext{1}{Astronomy Section, Nagoya City Science Museum, 2-17-1 Sakae, Naka-ku, Nagoya, Aichi 460-0008, Japan}
\altaffiltext{2}{Department of Physics, Graduate School of Science, Nagoya University, Furo-cho, Chikusa-ku, Nagoya, Aichi 464-8602, Japan}
 \author{Yoshiaki Sofue\altaffilmark{3}}%
 \altaffiltext{3}{Institute of Astronomy, The University of Tokyo, 2-21-1 Osawa, Mitaka, Tokyo 181-0015, Japan}
 
\email{kohno@nagoya-p.jp}
\email{mikito.kohno@gmail.com}


\maketitle

\begin{abstract}
We performed correlation analyses between the $^{12}$CO and $^{13}$CO $J=$1-0 line intensities in order to derive {the variability of} the CO-to-H$_2$ conversion factor ($X_{\rm CO, {iso}}$) in the central molecular zone (CMZ) of our Galaxy. 
New high-resolution $X_{\rm CO, {iso}}$ maps at a resolution of $\sim \timeform{30"}$ and the longitude-velocity diagram (LVD) at resolution $\sim \timeform{30"}\times 2$ \kms\ are presented using the $^{12}$CO and $^{13}$CO archival survey data obtained by the Nobeyama 45 m telescope.
{We revealed the variation of $X_{\rm CO, {iso}}$ in the CMZ within the range of $X_{\rm CO, {iso}} \sim (0.2-1.3) \times 10^{20}\ {\rm cm^{-2}\ (K\ km\ s^{-1})^{-1}}$, if we assume the normalization value of $0.59 \times 10^{20}\ {\rm cm^{-2}\ (K\ km\ s^{-1})^{-1}}$. 
The mean value is obtained as $X_{\rm CO, {iso}} = (0.48 \pm 0.15) \times 10^{20}\ {\rm cm^{-2}\ (K\ km\ s^{-1})^{-1}}$ in the CMZ of our Galaxy.}
\end{abstract}

\section{Introduction} 
The Central Molecular Zone (hereafter CMZ) is the concentration of molecular gas within $\sim 300$ pc in the Galactic Center (GC: \cite{1996ARA&A..34..645M,2023ASPC..534...83H}). 
The total molecular mass in the CMZ has been derived to be $\sim 2.3 \times 10^7\ M_{\odot}$ assuming the CO-to-H$_2$ conversion factor of $\Xg=0.51\times 10^{20}$  cm$^{-2}$ (\Kkms)$^{-1}$ \citep{2022MNRAS.516..907S}, but it directly depends on the assumed value of $\Xg$, which is still uncertain in the GC \citep{1998A&A...331..959D}.
Hence, the mass, the most fundamental parameter of the GC, is still uncertain by a factor of $\sim 2$ to 4.
The conversion factor is defined by
\begin{equation}
X_{\rm CO} = {N({\rm H_2})_{\rm 12X} \over I_{\rm ^{12}CO}} \ {\rm [cm^{-2}\ (K\ km\ s^{-1})^{-1}]},
\label{XCO}
\end{equation}
where $N({\rm H_2})_{\rm 12X}$ is the H$_2$ column density and $I_{\rm ^{12}CO}$ is the integrated intensity of the $^{12}$CO $J=$1-0 line. The standard value of the $\Xgeneral$ factor is $2 \times 10^{20}$ ${\rm cm^{-2} (K\ km\ s^{-1})^{-1}}$ with the uncertainty of $\pm 30\%$ in the Galactic disk \citep{2013ARA&A..51..207B}.

In the GC, $\Xg$ is suggested to be significantly smaller than the local value due to the higher metallicity \citep{arimoto+1996}, decreasing by a factor of $\sim 3$--10 \citep{2013ARA&A..51..207B}. 
Lower $\Xgeneral$ factors have been also pointed out comparison of the CO luminosity with the $\gamma$ ray brightness \citep{1985A&A...143..267B,2004A&A...422L..47S}, analysis of the infrared dust emission \citep{1995ApJ...452..262S,2014A&A...566A.120S}, and Virial mass \citep{1998ApJ...493..730O}. 

Recently, Sofue \& Kohno (2020, hereafter: Paper I) and Kohno \& Sofue (2024, hereafter: Paper II) revealed the variability of the $\Xgeneral$ factor in the Galactic giant molecular clouds (GMC). These papers present the high-resolution ($\sim 15\arcsec$) $\Xgeneral$ maps using CO isotopologues obtained by the FUGIN CO survey data \citep{2017PASJ...69...78U,2019PASJ...71S...2T}.

In this paper, we aim first at deriving the $\Xgeneral$ {variability} in the GC at higher accuracy and at presenting detailed distributions of $\Xg$ in longitude-latitude 
 $(l,b)$ and longitude-velocity $(l,V_{\rm LSR})$ planes at high angular and velocity resolutions using the CO-line data from the Nobeyama 45-m radio telescope. 

\section{Data} 
We used the high-resolution CO $J=$1-0 archival data of the Galactic center obtained by the Nobeyama 45 m telescope \citep{2019PASJ...71S..19T}\footnote{\url{https://www.nro.nao.ac.jp/~nro45mrt/html/results/data.html}}.
The $^{12}$CO and $^{13}$CO $J=$ 1-0 data were obtained by the multi-beam receiver of BEARS (25-BEam Array Receiver System: \cite{2000SPIE.4015..237S,2000SPIE.4015..614Y}) and FOREST (FOur beam REceiver System on the 45-m Telescope \cite{2016SPIE.9914E..1ZM,2019PASJ...71S..17N}) installed in the Nobeyama 45 m telescope. {The full width half-maximum beam size is approximately \timeform{15"} at 115 GHz and 110 GHz. 
The intensity is calibrated to the radiation temperature ($T_R^*$: \cite{1981ApJ...250..341K}) comparing with the previous CO survey data observed by the Nobeyama 45 m telescope \citep{1998ApJS..118..455O}. 
Here, $T_R^*$ corresponds to the brightness temperature $\Tb$ of a spatially resolved (extended) source. The final data used in this paper has a grid spacing with $(l,b,v) = (\timeform{7.5"},\timeform{7.5"}, 2\ {\rm km\ s^{-1}})$. 
We smoothed the archival data to $\sim \timeform{30"}$ using the kernel Gaussian function to improve the signal-to-noise ratio.} The root-mean-square (r.m.s) noise levels of the final cube data are $\sim 0.4 $ K and $\sim 0.1$ K for $^{12}$CO and $^{13}$CO $J=$1-0, respectively.

\section{Methods}
In this study, we calculated the column densities of molecular hydrogen using the integrated intensity of $^{12}$CO and $^{13}$CO $J=$1-0. The H$_2$ column density is taken from the $^{12}$CO integrated intensity by assuming the $\Xgeneral$ factor from the equation (\ref{XCO}).
We used the conversion factor of ${\Xgeneral} = 0.59 \times 10^{20}\ {\rm cm^{-2}\ (K\ km\ s^{-1})^{-1}}$ \citep{arimoto+1996} currently obtained for the Galactic Center region as the normalization value in this article. 
This value has been obtained by extrapolating the $\Xgeneral$ values determined in the disc to the GC by assuming a scale radius of the Galactic disc, $r_e = 6.2$ kpc, and the distance of the GC from the Sun, $R_0 = 8.0$ kpc \citep{2019ApJ...885..131R,2020PASJ...72...50V}. 

Independent of the conversion factor, we derived the H$_2$ column density in the CMZ by the local thermal equilibrium (LTE) method described in Paper I, II, and \citet{2008ApJ...679..481P}. 
The brightness temperature ($T_{\rm B}$) of the CO line intensity with the excitation temperature ($T_{\rm ex}$) and the optical depth ($\tau$) is given by

\begin{equation}
T_{\rm B} = T_0
\left(\frac{1}{e^{T_0/\Tex}-1} - \frac{1}{e^{T_0/T_{\rm bg} }-1}\right)
\left(1-e^{-\tau}\right)\ [{\rm K}],
\label{Tb}
\end{equation} 
where {$T_{\rm bg}=2.73$} K is the temperature of the cosmic-microwave background radiation.
$T_0=h \nu/k$ is the Planck temperature with $h$, $\nu$, and $k$ being the Planck constant, rest frequency, and Boltzman constant, respectively. 

If we assume that the $^{12}$CO line is optically thick, the excitation temperature is given by

\begin{equation}
\Tex=T_{0}^{115} \bigg/ {\rm ln} \( 1+\frac{T_{0}^{115}}{\Tbcotw_{\rm max}+{0.836}}  \)\ [{\rm K}],
\label{Tex}
\end{equation}
where $\Tbcotw_{\rm max}$ and $T_{0}^{115}={5.53}$ K correspond to the $^{12}$CO peak {brightness temperature} and the Planck temperature at the rest frequency of $^{12}$CO $J=$1-0, respectively. {In the longitude-latitude ($l-b$) and longitude-velocity ($l-v$) space, $T_{\rm ex}$ at each pixel is calculated using the peak value along the velocity-axis and latitude-axis, respectively.} We assume that $T_{\rm ex}$ is equal in the $^{12}$CO and $^{13}$CO line emissions, and express the optical depth as
\be
\tau(^{13}{\rm CO})=-{\rm ln} 
\(1-\frac{\Tbcoth_{\rm max}/T_{0}^{110} }{(e^{T_{0}^{110}/\Tex}-1)^{-1}- {0.168}} \),
\ee
where $\Tbcoth_{\rm max}$ and $T_{0}^{110}={5.29}$ K represent the $^{13}$CO peak {brightness temperature} and the Planck temperature at the rest frequency of $^{13}$CO $J=$1-0, respectively. 
The $^{13}$CO column density is given by
\be
\Ncoth=3.0\times 10^{14} ~ \frac{\tau}{1-e^{-\tau}}
\frac{I_{\rm ^{13}CO}}{ 1-e^{-T_{0}^{110}/\Tex}}\ [{\rm cm^{-2}}],
\ee
where $I_{\rm ^{13}CO}$ is the $^{13}$CO integrated intensity.
Then, we convert $\Ncoth$ to the H$_2$ column density using the abundance ratio of H$_2$ to $^{13}$CO molecules given by
\be
N({\rm H_2})_{\rm 13L}=Y_{\rm GC ^{13}CO} \Ncoth\ [{\rm cm^{-2}}].
\label{YN}
\ee
Here, $Y_{\rm GC ^{13}CO}$ is adopted as $\sim 1\times 10^6$ in Sagittarius B2 \citep{1989ApJ...337..704L}.

\section{Results}
\begin{figure}
\begin{center} 
\includegraphics[width=8.5cm]{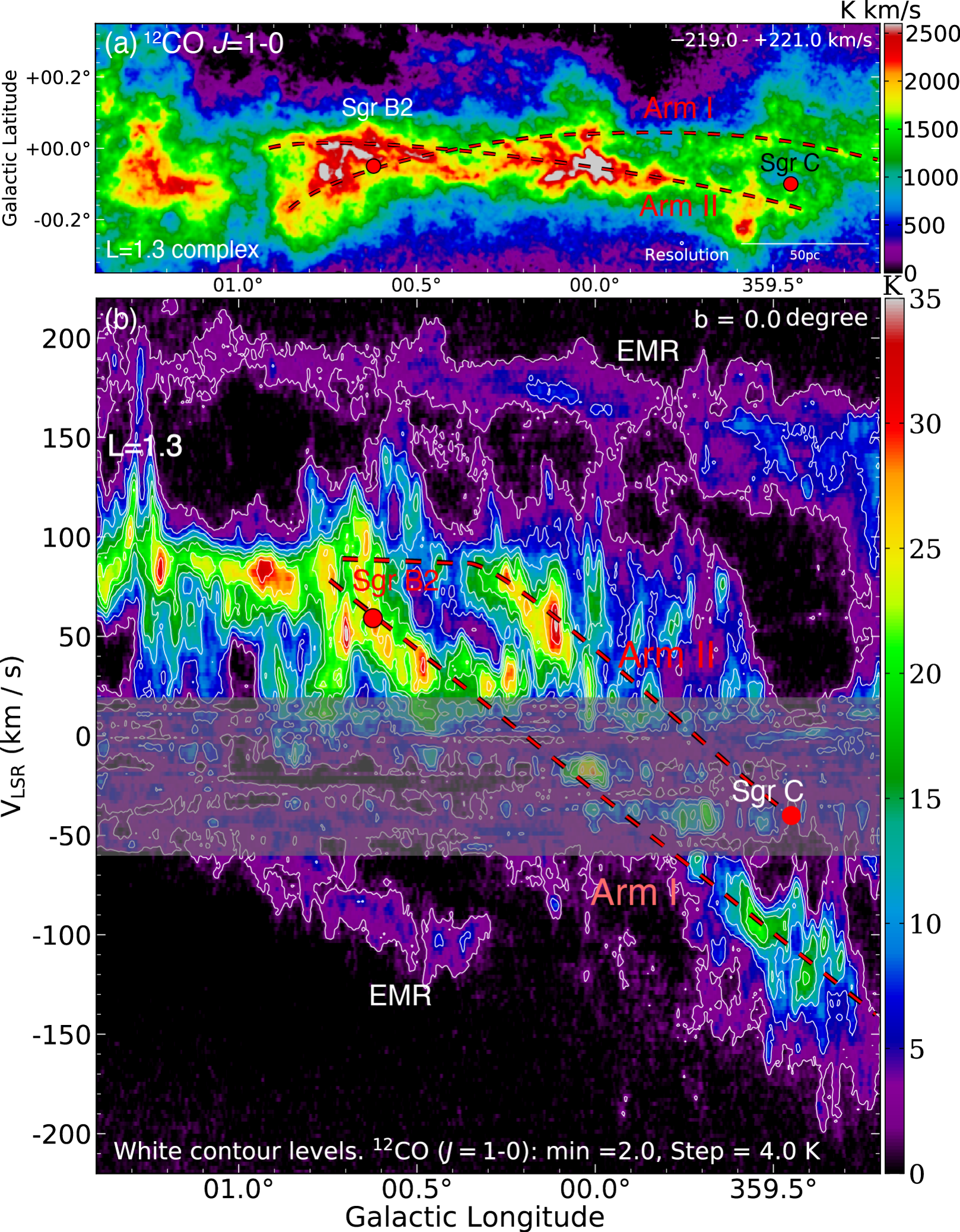}
\end{center}
\caption{(a) The integrated intensity map of $^{12}$CO $J=$1-0  obtained by the Nobeyama 45 m telescope \citep{2019PASJ...71S..19T}. The integrated velocity range is from $-219$ \kms to 221 \kms. (b) The longitude-velocity diagram of $^{12}$CO at $b=0 \deg$. Arm I and Arm II are pointed in red {dashed} lines, that is spiral arms around Sgr A \citep{1995PASJ...47..527S}. The lowest white contour level and intervals of the panel are 2.0 and 4.0 K. {The shadow area from $-60$ to $+20$ \kms is contaminated by the foreground spiral arms (e.g., \cite{2001ApJ...562..348O,2006PASJ...58..335S,2016ApJ...823...77R})}. The red points show the positions of Sgr B2 and Sgr C.} 
\label{GCmap}
\end{figure}
\begin{figure}
\begin{center}        
\includegraphics[width=8cm]{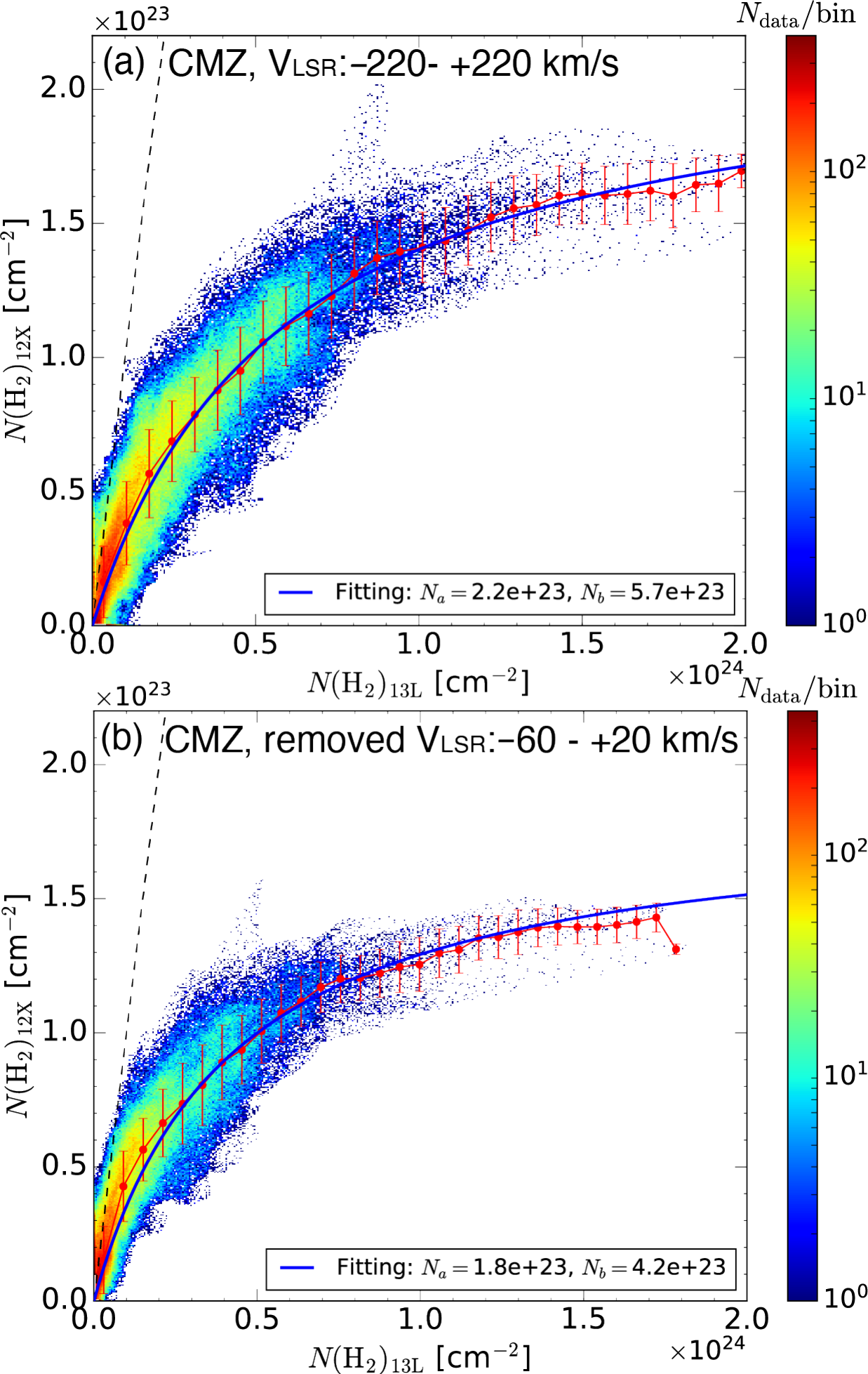}
\end{center}
\caption{(a) Scatter plots between $N({\rm H_2})_{\rm 13L}$ and $N({\rm H_2})_{\rm 12X}$ of the CMZ.
Red points show the averaged values of each bin, and the error bars are standard deviations of $N({\rm H_2})_{\rm 12X}$. Blue curves indicate the fitting results of scatter plots. The black dashed lines indicate the linear relation of $N({\rm H_2})_{\rm 12X}=N({\rm H_2})_{\rm 13L}$. (b) Same as (a), but for {contaminated velocity range} at $-60 \le \vlsr < +20$ \kms are removed. The color bars show the number of data points in a bin ($N_{\rm data}/{\rm bin}$).} 
\label{CMZscatter}
\end{figure}

\subsection{CO spatial and velocity distributions in the CMZ}
Figure \ref{GCmap} (a) shows the integrated intensity map of $^{12}$CO $J=$1-0. The integrated velocity range is from $-219$ \kms to $+221$ \kms. $^{12}$CO has a peak at the GMC complex of Sagittarius B2 (Sgr B2), Sagittarius C (Sgr C), and $l=1\deg.3$ complex. Molecular gas shows $\sim 50$ pc distributions of each GMC complex. 
Figure \ref{GCmap}(b) shows the longitude-velocity diagram (LVD) of $^{12}$CO $J=$ 1-0 cutting at $b=\timeform{0D}$.
The red {dashed} lines show Arm I and Arm II of the spiral arms around Sgr A \citep{1995PASJ...47..527S}. 
The CMZ is distributed from $-200$ \kms to $+200$ \kms on the velocity space. 
We can also find the Expanding Molecular Ring (EMR) with the non-circular gas motion around the GC \citep{1972NPhS..238..105K,1974PASJ...26..117K,1972ApJ...175L.127S,1995PASJ...47..551S}.
{The shadowed area at the velocity range from $-60$ \kms to $+20$ \kms is contaminated by the GC's foreground spiral arms \citep{1998ApJS..118..455O,2006PASJ...58..335S}.} 
\begin{figure}
\begin{center}        
\includegraphics[width=8.5cm]{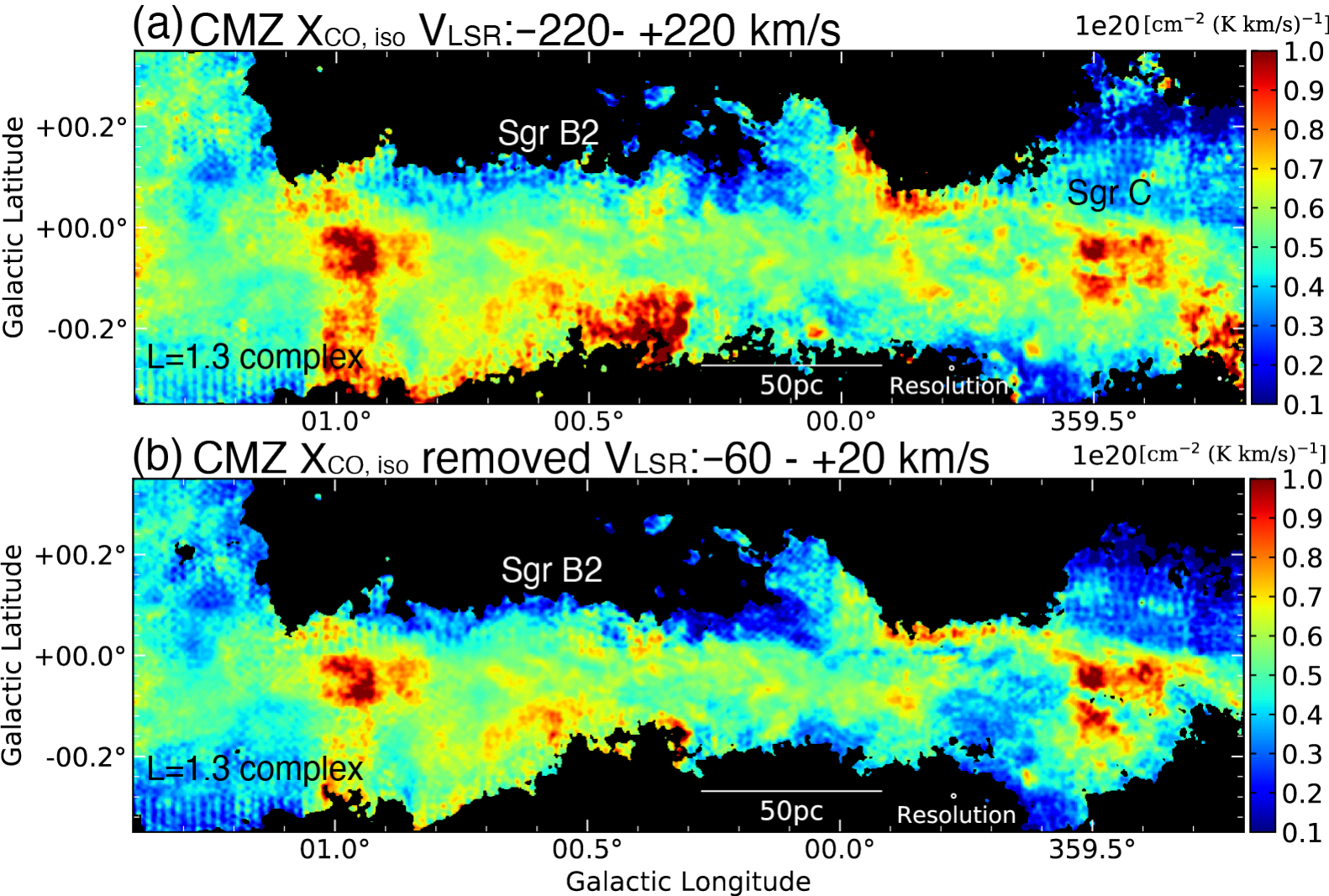}\\
\end{center}
\caption{(a) Spatial distributions of $X_{\rm CO, iso}$ in the CMZ. The map is presented of $I_{\rm ^{12}CO} > 500$ K \kms. 
{(b) Same, but {contaminated velocity range} at $-60 \le \vlsr <+20$ \kms are removed.}
} 
\label{CMZXCOmap}
\end{figure}
\begin{table*}
\tbl{Fitting results and $<X_{\rm CO, {iso}}>$ taken from the correlation between $N({\rm H_2})_{\rm 13L}$ and $N({\rm H_2})_{\rm 12X}$.}{
\begin{tabular}{cccccccc} 
\hline 
\hline   
Name & $l$ & $b$ &$V_{\rm LSR}$ & $N_{\rm a}$ & $N_{\rm b}$ & $<X_{\rm CO, iso}>$ & C.C\\
&[deg]&[deg]& [\kms] &{[$ 10^{23}$ cm$^{-2}$]} & {[$ 10^{23}$ cm$^{-2}$]} & {[$ 10^{20}$ ${\rm cm^{-2} (K\ km\ s^{-1})^{-1}}$]}& \\
(1) & (2) & (3) & (4)  & (5)& (6) &(7) &(8)   \\
\hline
CMZ (all)  & --- & --- & [$-220$, $220$] & 2.2 & 5.7  &  $0.54 \pm 0.16$ & 0.89\\
\multirow{2}{*}{CMZ$^{[1]}$} & \multirow{2}{*}{---} & \multirow{2}{*}{---} & [$-220$, $-60$] & \multirow{2}{*}{1.8} & \multirow{2}{*}{4.2}  &  \multirow{2}{*}{$0.48 \pm 0.15$} & \multirow{2}{*}{0.87}\\
&&&[$20$, $220$] &&&&\\
Sgr B2 & $0.62$ & $-0.05$ & [$20$, $150$] & 1.9 & 4.7 & $0.54\pm 0.14$ & 0.92\\
Sgr C & $-0.55$ & $-0.10$ & [$-90$, $-30$] & 0.46 & 1.1 & $0.56\pm 0.16$ & 0.87\\
$l=$\timeform{1.3D} & 1.3 & $-0.1$ & [$20$, $150$] & 1.5  & 2.3 & $0.59\pm 0.11$ & 0.92\\
\hline   
\hline
\end{tabular}}
\label{tab_region}
[1] {The velociy components from $-60$ \kms to $20$ \kms were removed.}
Columns: (1) Region names,  (2) Galactic longitude, (3) Galactic latitude, (4) Integrated velocity range, (5),(6) The fitting parameters of the non-linear relation. (7) The mean value of $X_{\rm CO, {iso}}$ in the each region. (8) Correlation coefficient. The errors of mean value are adopted as the {standard deviation} in all pixels. $<X_{\rm CO, iso}>$ of the individual regions are derived by the data points of $I_{\rm ^{12}CO} > 200$ K \kms.
\end{table*} 

\subsection{$X_{\rm CO, {iso}}$ maps} 
Figure \ref{CMZscatter} shows the scatter plots between $N({\rm H_2})_{\rm 13L}$ and $N({\rm H_2})_{\rm 12X}$ in the CMZ. 
Black dashed line indicates the linear relation of $N({\rm H_2})_{\rm 12X}=N({\rm H_2})_{\rm 13L}$. $N({\rm H_2})_{\rm 12X}$ shows saturation at high $N({\rm H_2})_{\rm 13L}$, and the nonlinear relation. 
Then, we carried out the empirical curve fitting using the relation of
\be
N({\rm H_2})_{\rm 12X}  = {N_{\rm a} N({\rm H_2})_{\rm 13L} \over N({\rm H_2})_{\rm 13L} + N_{\rm b}}\ [{\rm cm}^{-2}],
\label{eq:fitting}
\ee
where $N_{\rm a}$ and $N_{\rm b}$ [both in \uNHtwo] are fitting parameters of the nonlinear relation having the same dimension as the column density.
From the equation \ref{XCO} and \ref{eq:fitting}, we calculated the $\Xgeneral$ factor (hereafter $X_{\rm CO, iso}$) of each pixel using the following relation of
\be 
X_{\rm CO, {iso}} = \left({N_{\rm a} N({\rm H_2})_{\rm 13L} \over N({\rm H_2})_{\rm 13L} + N_{\rm b}}\right) \bigg/ I_{\rm ^{12}CO}\ {\rm [cm^{-2} (K\ km\ s^{-1})^{-1}]}.
\ee
Figure \ref{CMZXCOmap}(a) and \ref{CMZXCOmap}(b) show the $X_{\rm CO, iso}$ maps of the CMZ with the integrated velocity range from $-220$ to $+220$ \kms and removed velocity range from $-60$ to $+20$ \kms, respectively. $X_{\rm CO, {iso}}$ in the GMZ shows the variation of $(0.2-1.0) \times 10^{20}$ cm$^{-2}$ (K km s$^{-1}$)$^{-1}$. 
We obtained the mean value as $X_{\rm CO, {iso}} = (0.54 \pm 0.16) \times 10^{20}$ and $X_{\rm CO, {iso}} = (0.48 \pm 0.15) \times 10^{20}$ ${\rm cm^{-2} (K\ km\ s^{-1})^{-1}}$ in Figure \ref{CMZXCOmap}(a) and \ref{CMZXCOmap}(b), respectively. 
The mean $X_{\rm CO, iso}$ removed from the {contaminated} components is lower than all integrated velocity ranges. 
We also performed the curve fitting of the correlation between $N_{\rm a}$ and $N_{\rm b}$ of the individual regions in the CMZ. As a result, the mean $X_{\rm CO, iso}$ is obtained in Sgr B2, Sgr C, and the $L=\timeform{1.3D}$ complex.
Table \ref{tab_region} presents the fitting parameters and $X_{\rm CO, iso}$ of our analysis. 

\subsection{The $X_{\rm CO, {iso}}$ longitude-velocity diagram}
\begin{figure}
\begin{center}        
\includegraphics[width=8cm]{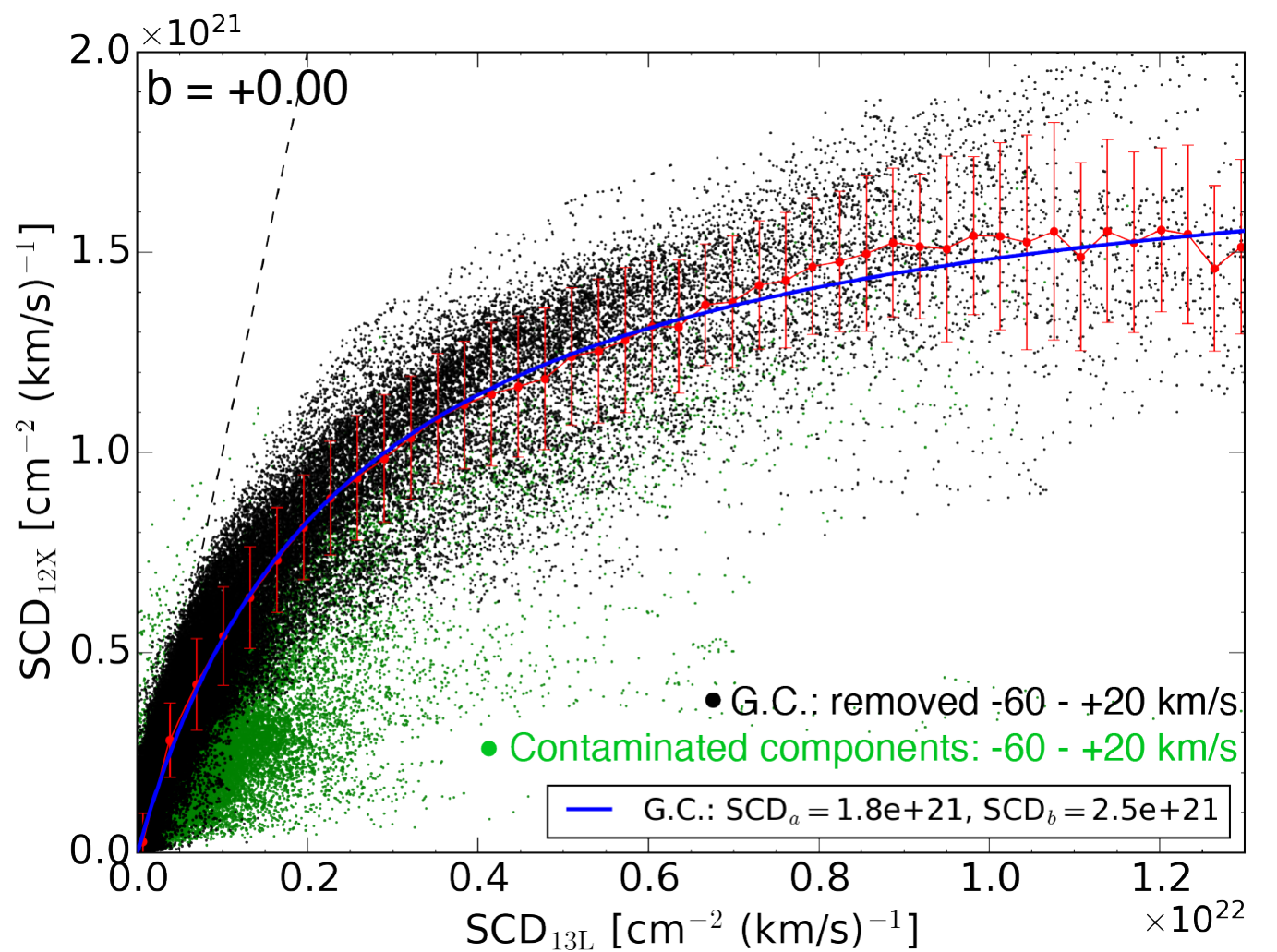}
\end{center}
\caption{Scatter plots between ${\rm SCD}_{\rm 13L}$ and ${\rm SCD}_{\rm 12X}$ on the longitude-velocity space at $b=\timeform{0.0D}$. 
{Black points show the $-220$ \kms $< V_{\rm LSR} < -60$ \kms and $+20$ \kms $< V_{\rm LSR} < +220$ \kms component. Green points present the $-60$ \kms $< V_{\rm LSR} < +20$ \kms component.} Red points show the averaged values of each bin, and the error bars are standard deviations of $N({\rm H_2})_{\rm 12X}$. The blue curve indicates the fitting result of the Galactic center components. The black dashed lines indicate the linear relation of ${\rm SCD}_{\rm 12X}={\rm SCD}_{\rm 13L}$.} 
\label{lvscatter}
\end{figure}
We made the $X_{\rm CO, iso}$ LVD of the CMZ at the Galactic plane using the correlation of spectral column densities (SCD) obtained by the $^{12}$CO and $^{13}$CO intensity described as Paper I and Paper II.
The ${\rm SCD}_{\rm 12X}$ and ${\rm SCD}_{\rm 13L}$ is given by
\begin{eqnarray}
{\rm SCD}_{\rm 12X} &=& \frac{d N({\rm H_2})_{\rm 12X}}{dv} \nonumber\\
&=& \Xgeneral \Tbcotw\ {\rm [cm^{-2} (km\ s^{-1})^{-1}]}
\label{SCD12X}
\end{eqnarray}
and
\begin{eqnarray}
{\rm SCD}_{\rm 13L} &=& \frac{d N({\rm H_2})_{\rm 13L}}{dv}\ {\rm [cm^{-2} (km\ s^{-1})^{-1}]} \nonumber \\ 
&=& 3.0\times 10^{14}~ \frac{\tau}{1-e^{-\tau}} \frac{\Ycoth \Tbcoth}{ 1-e^{-T_{0}^{110}/\Tex}},
\label{SCD13L}
\end{eqnarray}
respectively. 
${\rm SCD}_{\rm 12X}$ and ${\rm SCD}_{\rm 13L}$ are the H$_2$ column densities at assuming a normalization value of ${\Xgeneral} = 0.59 \times 10^{20}$ cm$^{-2}$ (K km s$^{-1}$)$^{-1}$ \citep{arimoto+1996} and $Y_{\rm GC ^{13}CO} = 1\times 10^6$ \citep{1989ApJ...337..704L}, respectively.
Figure \ref{lvscatter} shows scatter plots between ${\rm SCD}_{\rm 12X}$ and ${\rm SCD}_{\rm 13L}$ in the longitude-velocity space at the Galactic plane ($b=\timeform{0.0D}$). {Scatter plots on the longitude-velocity space at the different latitudes are presented in Figures \ref{app+0.30}a-\ref{app-0.3}a of the Appendix.}
{Green data points indicate from $-60$ \kms to $+20$ \kms contaminated by the foreground disk components. Black data points show the CMZ components with the removed from $-60$ \kms to $+20$ \kms components.} 
The SCD$_{\rm 12X}$ saturation level of the GC is higher than {contaminated} components. We suggest that this result corresponds to the difference of the $^{12}$CO and $^{13}$CO ratio between the CMZ and the Galactic disk components reported in previous works (e.g., Figure 10 in \cite{2023PASJ...75..970E}). 
Here, we performed the curve fitting to the GC components using the function of
\be
{\rm SCD}_{\rm 12X}  = {{\rm SCD}_{\rm a} {\rm SCD}_{\rm 13L} \over {\rm SCD}_{\rm 13L} + {\rm SCD}_{\rm b}}\ {\rm [cm^{-2} (km\ s^{-1})^{-1}]},
\label{eq:fittinglv}
\ee
where ${\rm SCD}_{\rm a}$ and ${\rm SCD}_{\rm b}$ correspond to the fitting parameters.
The result of fitting parameters are obtained as ${\rm SCD}_{\rm a}=1.8\times 10^{21}\ {\rm cm^{-2} (km\ s^{-1})^{-1}}$ and ${\rm SCD}_{\rm b}=2.5\times 10^{21}\ {\rm cm^{-2} (km\ s^{-1})^{-1}}$. Then, $X_{\rm CO, {iso}}$ in the LVD is given by
\begin{eqnarray}
X_{\rm CO, {iso}} = \left({{\rm SCD}_{\rm a} {\rm SCD}_{\rm 13L} \over {\rm SCD}_{\rm 13L} + {\rm SCD}_{\rm b}}\right)  \bigg/ \Tbcotw &  \nonumber\\
{\rm [cm^{-2} (K\ km\ s^{-1})^{-1}]}&.
\end{eqnarray}


\begin{table}
{
\tbl{$X_{\rm CO, iso}$ on the LVD.}{
\begin{tabular}{cccccc} 
\hline 
\hline   
Name &$V_{\rm LSR}$ & $X_{\rm CO, iso}$ \\
& [\kms] & {[$ 10^{20}$ ${\rm cm^{-2} (K\ km\ s^{-1})^{-1}}$]}& \\
\hline
{Arm I} &---& $\sim 0.5$ -- 1.0\\
{Arm II} &---& $\sim 0.4$ -- 0.8\\
\multirow{2}{*}{EMR} &[$150$, $200$]& \multirow{2}{*}{$\sim 0.2$ -- 1.3}\\
&[$ -200$, $-100$]&\\
\hline   
\hline
\end{tabular}}
\label{tab_lv}
} 
\end{table}
\begin{figure}
\begin{center}        
\includegraphics[width=8.4cm]{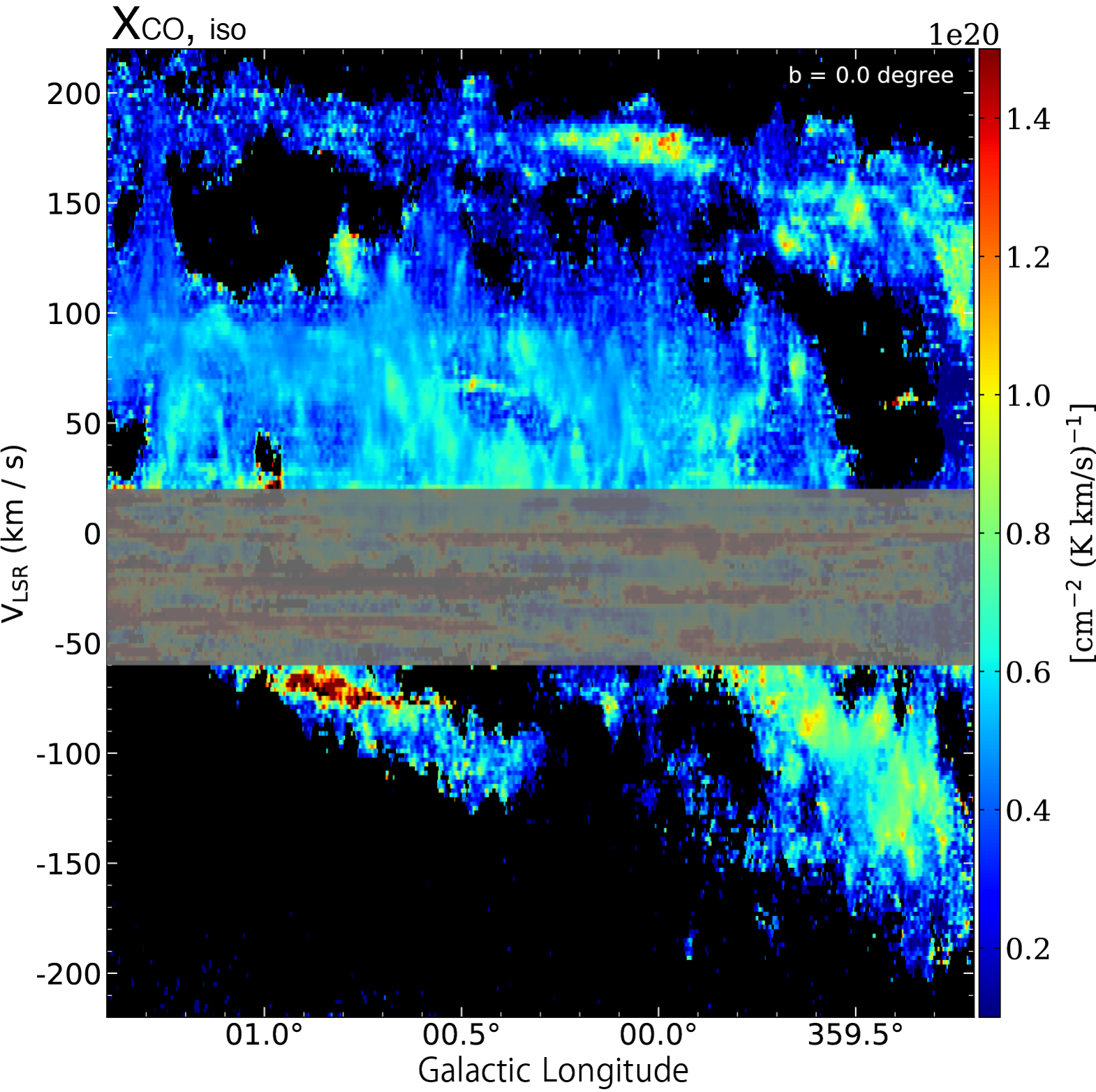}
\end{center}
\caption{{Longitude-velocity distributions of $X_{\rm CO, {iso}}$ in the CMZ at $b=\timeform{0D}$. The map is presented for $T_{\rm B} ({\rm ^{12}CO)} > 1$ K. {The masked area shows the contaminated velocity range by the foreground disk components.} }}
\label{lvXCO}
\end{figure}

{Figure \ref{lvXCO}} shows the $X_{\rm CO, iso}$ map on the LVD at $b=\timeform{0.0D}$. {$X_{\rm CO, iso}$ maps on the LVDs at the different latitudes are presented in Figures \ref{app+0.30}b-\ref{app-0.3}b of the Appendix.}
The masked area presents the contaminated velocity range from the foreground disk components. 
This velocity range exists the CMZ gas and foreground disk gas in the same pixel of the data (see Figures \ref{12COch} and \ref{13COch} in the Appendix). In the optically thick $^{12}$CO line, the emission from the CMZ is completely obscured and we only see the foreground gas. On the other hand, in the optically thin $^{13}$CO line, we practically see the sum of the CMZ emission and foreground emission. In such a situation, the $^{12}$CO and $^{13}$CO lines trace totally different components.
In this paper, we therefore focus on the variation of $X_{\rm CO, iso}$ in the CMZ components, which are removed the contaminated velocity range on the LVD.
$X_{\rm CO, iso}$ has the variability of $X_{\rm CO, iso} \sim (0.2-1.3) \times 10^{20}{\rm cm^{-2} (K\ km\ s^{-1})^{-1}}$ in the CMZ component on the LVD. We summarize the $X_{\rm CO, iso}$ values on the LVD of Arm I, Arm II, and EMR in Table \ref{tab_lv}.

\section{Discussion}
{We investigated the radial variation of $X_{\rm CO, iso}$ based on the results of Paper II.
Figure \ref{gradient} shows the plot of $X_{\rm CO, iso}$ as a function of the Galactocentric distance. $X_{\rm CO, iso}$ of the GMCs in the Galactic disk were obtained by Paper II assuming the normalization value of $\Xgeneral =2 \times 10^{20}\ {\rm cm^{-2} (K\ km\ s^{-1})^{-1}}$\citep{2013ARA&A..51..207B}.
In this paper, we assumed the normalization in the GC of $\Xgeneral=0.59\times 10^{20}\ {\rm cm^{-2} (K\ km\ s^{-1})^{-1}}$\citep{arimoto+1996}.
Since the obtained value is proportional to this normalization factor if adopting a value of $2.0\times 10^{20}\ {\rm cm^{-2} (K\ km\ s^{-1})^{-1}}$ of local molecular clouds, the whole results in our analysis must be multiplied by a factor of 3.39. 
The pink and yellow triangle in Figure \ref{gradient} shows $X_{\rm CO, iso}$ in the GC assuming the normalization $\Xgeneral$ of $2.0\times 10^{20}\ {\rm cm^{-2} (K\ km\ s^{-1})^{-1}}$ and $0.59\times 10^{20}\ {\rm cm^{-2} (K\ km\ s^{-1})^{-1}}$, respectively.
\begin{figure}
\begin{center}        
\includegraphics[width=8cm]{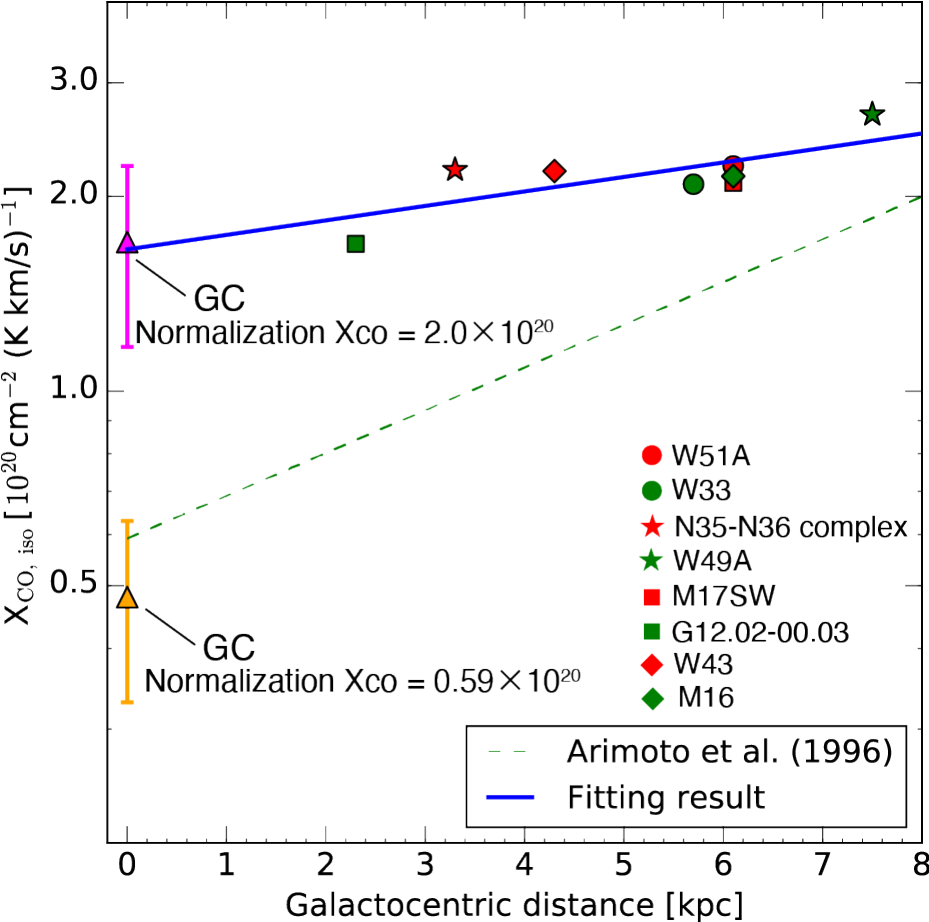}
\end{center}
\caption{Radial gradient of $X_{\rm CO, iso}$ from the distance of the GC. {The pink and yellow triangle shows $X_{\rm CO, iso}$ in the GC assuming the normalization $\Xgeneral$ of $2.0\times 10^{20}\ {\rm cm^{-2} (K\ km\ s^{-1})^{-1}}$ and $0.59\times 10^{20}\ {\rm cm^{-2} (K\ km\ s^{-1})^{-1}}$, respectively. The vertical error bars of the GC indicate the standard deviation in the $X_{\rm CO, iso}$ map.}
Other symbols present the individual GMCs of W51A, W33, N35-N36 complex, W49A, M17SW, G012.02-00.03, W43, and M16 obtained by Paper II.
{The blue line shows the fitting result to individual GMCs and GC assuming the normalization $\Xgeneral = 2 \times 10^{20}\ {\rm cm^{-2} (K\ km\ s^{-1})^{-1}}$ by an exponential function. The green dashed line indicates the relation from \citet{arimoto+1996}.} 
} 
\label{gradient}
\end{figure}

\begin{table}
\tbl{Comparison with the previous studies of the $\Xgeneral$ factor in the Galactic Center region. }{
\begin{tabular}{cccccc} 
\hline\hline
Method  &$X_{\rm CO}$ & References \\
 &{$[ 10^{20}$ ${\rm cm^{-2} (K\ km\ s^{-1})^{-1}}]$} & &\\
\hline
Isotopologues  & {0.2--1.3$^{\dag}$} & {This study}\\
  & {0.7--4.4$^{*}$} & {This study}\\
\hline
Virial  & 0.24 & [1]\\
Gamma-rays  & 0.3--0.6 & [2]\\
Extinction  & 0.3--0.6 & [3]\\
Dust  & 0.2--0.7 & [4] \\
Dust and HI  & $0.7\pm0.1$ & [5] \\
\hline
\hline
\end{tabular}}
\label{tab2}
We assumed the normalization $\Xgeneral$ value of $^\dag 0.59 \times 10^{20}$ ${\rm cm^{-2} (K\ km\ s^{-1})^{-1}}$ and $^* 2.0 \times 10^{20}$ ${\rm cm^{-2} (K\ km\ s^{-1})^{-1}}$. References [1] \citet{1998ApJ...493..730O},[2]\citet{2012ApJ...750....3A} [3] \citet{2014A&A...566A.120S},[4] \citet{1995ApJ...452..262S}, [5] \citet{2010PASJ...62.1307T}
\end{table} 
{Finally, we compared $X_{\rm CO, iso}$ in the CMZ with previous studies obtained by other methods to validate our assumption of the normalization value. The $\Xgeneral$ in the GC were taken from comparing the CO luminosity with the virial mass, gamma-ray brightness, extinction, infrared dust emission, and HI emission. 
Table \ref{tab2} summarizes the Galactic Center $\Xgeneral$ values.
Previous studies of $\Xgeneral$ in the GC are in the range of $(0.2-0.7) \times 10^{20}\ {\rm cm^{-2} (K\ km\ s^{-1})^{-1}}$, which is lower than the standard value of $2 \times 10^{20}\ {\rm cm^{-2} (K\ km\ s^{-1})^{-1}}$ in the Milky Way disk \citep{2013ARA&A..51..207B}. 
Our results using CO isotopologues assuming the normalization value of $\Xgeneral=0.59 \times 10^{20}\ {\rm cm^{-2} (K\ km\ s^{-1})^{-1}}$ shows the variation of {$X_{\rm CO, iso}\sim (0.2-1.3) \times 10^{20}\ {\rm cm^{-2} (K\ km\ s^{-1})^{-1}}$} in the GC, and its value is consistent with these previous works. 
These points show our assumption of the normalization $\Xgeneral$ of $0.59 \times 10^{20}\ {\rm cm^{-2} (K\ km\ s^{-1})^{-1}}$ based on \citet{arimoto+1996} is valid in this paper, whereas the fitting result shown in the blue line is consistent with the pink triangle obtained by the normaliation value of $2 \times 10^{20}\ {\rm cm^{-2} (K\ km\ s^{-1})^{-1}}$.
This inconsistency might be explained by the gradient between $X_{\rm CO, {iso}}$ in the CMZ and the Galactic disk component is a discontinuous function of the Galactocentric distance.}
{The absolute value of $X_{\rm CO, {iso}}$ is influenced by the initially-assumed $\Xgeneral$, while the relative variability of $X_{\rm CO, {iso}}$ is independet of the normalization value. We point out that our assumption does not change the difference of $X_{\rm CO, {iso}}$ on the longitude-latitude and longitude-velocity space.}

\section{Summary}
The conclusions of this paper are summarized as follows:
\begin{enumerate}
\item We investigated the correlation between the $^{12}$CO and $^{13}$CO $J=$1-0 integrated intensities and the brightness temperatures in the central molecular zone of our Galaxy using the archival CO survey data obtained by the Nobeyama 45 m telescope. The non-linear relation is revealed between them.
\item We performed the curve fitting of the non-linear relation between the $^{12}$CO and $^{13}$CO $J=$1-0 intensities to derive the variability of the CO-to-H$_2$ conversion factor ($X_{\rm CO, {iso}}$) in the CMZ.
\item The high-resolution $X_{\rm CO, {iso}}$ maps ($\sim \timeform{30"}$) and the longitude-velocity diagram are presented. {The $X_{\rm CO, {iso}}$ in the Galactic center region shows the variability of $X_{\rm CO, {iso}} \sim (0.2-1.3) \times 10^{20}\ {\rm cm^{-2}\ (K\ km\ s^{-1})^{-1}}$ if we assume the normalization value of $\Xgeneral=0.59 \times 10^{20}\ {\rm cm^{-2} (K\ km\ s^{-1})^{-1}}$}. 
\item {We obtained $X_{\rm CO,{iso}} = (0.48 \pm 0.15) \times 10^{20}$ ${\rm cm^{-2}}$ ${\rm (K\ km\ s^{-1})^{-1}}$ as the mean value in the CMZ of our Galaxy}.
\end{enumerate}


\section*{Acknowledgements}
{We are grateful to the anonymous referee for his/her thoughtful comments.}
The Nobeyama 45-m radio telescope is operated by the Nobeyama Radio Observatory.
We utilized the Python software package for astronomy \citep{2013A&A...558A..33A}.
{The authors are grateful to Prof. Tomoharu Oka and the radio astronomy group of Keio University for the archival GC survey data using the Nobeyama 45-m telescope.}

\appendix
\section*{{Velocity channel maps of the contaminated velocity range}}
{We present the $^{12}$CO and $^{13}$CO velocity channel maps contaminated by foreground disk components in Figures \ref{12COch} and \ref{13COch}, respectively. In this velocity range, the CMZ and foreground disk gas components coexist in the same data pixel. }
 \begin{figure*}
\begin{center}    
\includegraphics[width=18cm]{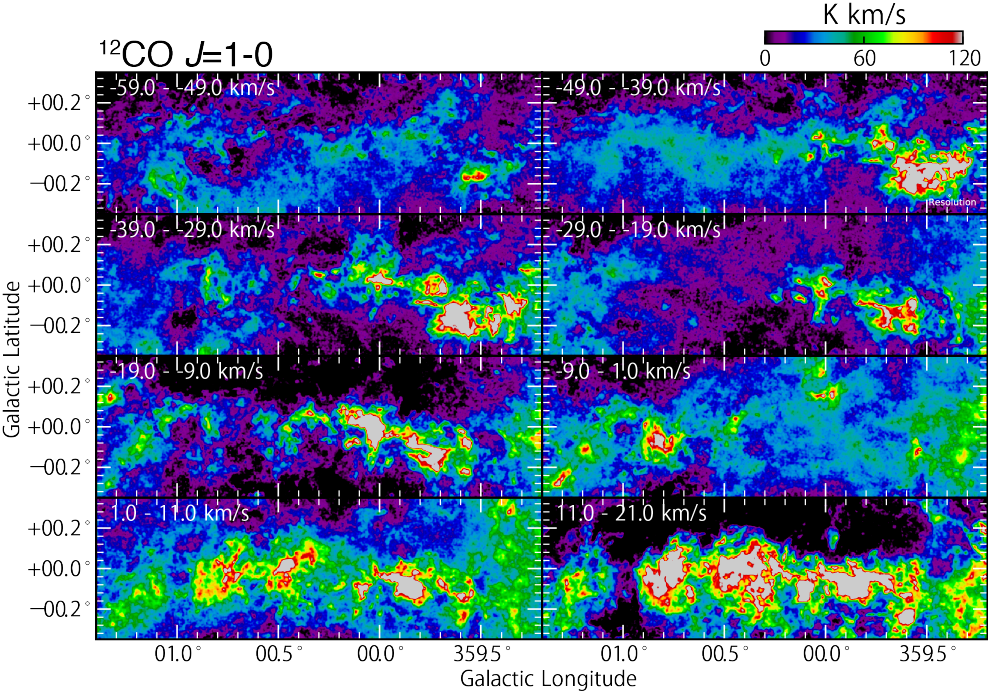}
\end{center}
\caption{{The $^{12}$CO $J=$1-0 channel map of the contaminated velocity range with the CMZ and foreground disk components.}}  
\label{12COch}
\end{figure*}

 \begin{figure*}
\begin{center}    
\includegraphics[width=18cm]{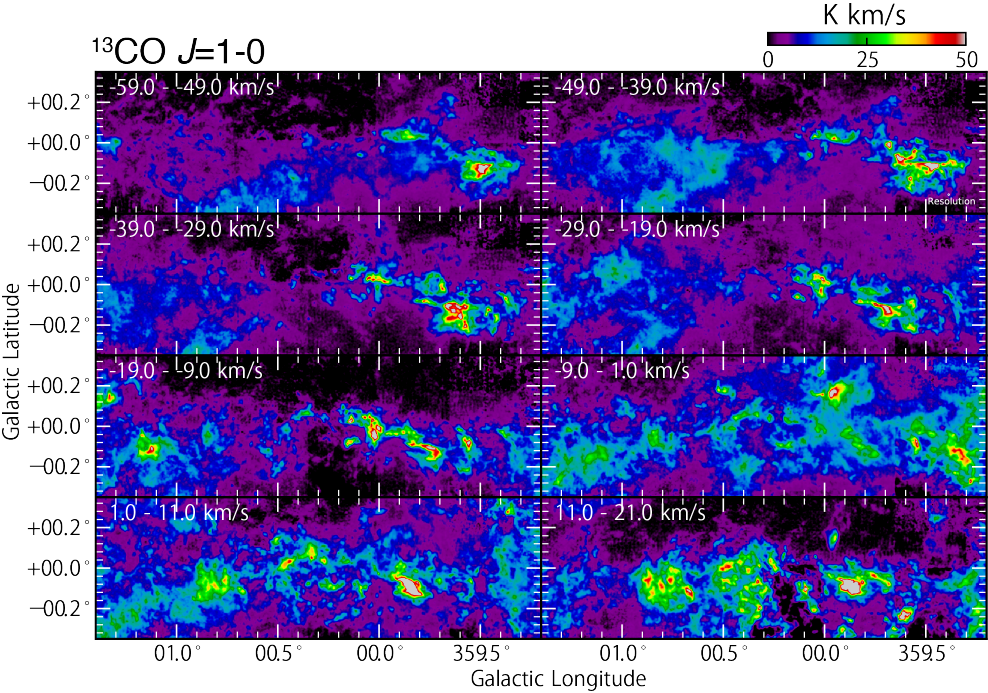}
\end{center}
\caption{{Same as Figure \ref{12COch}, but for $^{13}$CO $J=$1-0.}}  
\label{13COch}
\end{figure*}

\section*{{Scatter plots and $X_{\rm CO, iso}$ maps on the LVDs.}}
{We show scatter plots of spectral column densities and $X_{\rm CO, iso}$ maps on the LVDs from $b= \timeform{+0.30D}$ to $b= \timeform{-0.30D}$ in Figures \ref{app+0.30} - \ref{app-0.3}.}

 \begin{figure*}
\begin{center}    
\includegraphics[width=13.5cm]{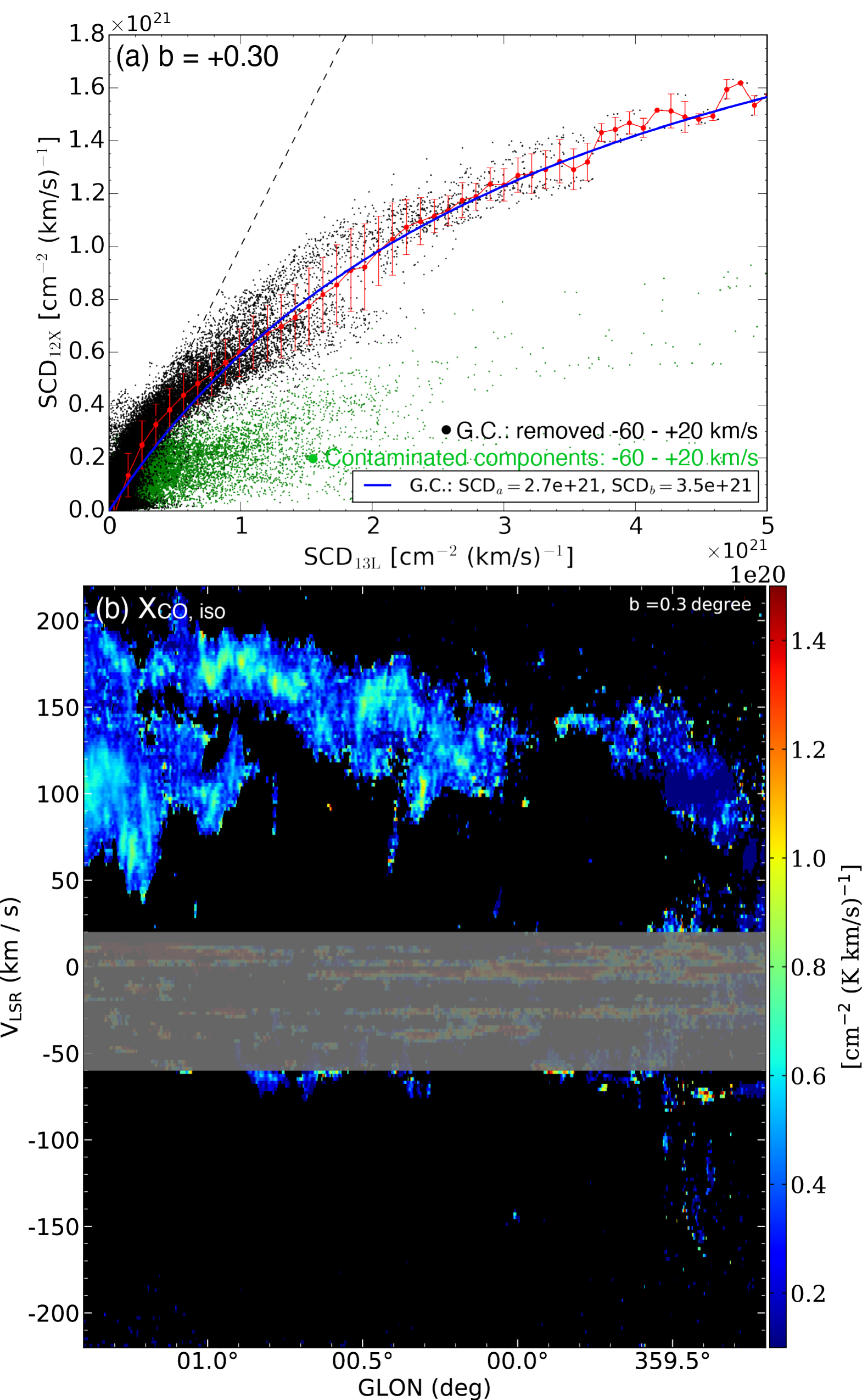}
\end{center}
\caption{{(a) Same as Figure \ref{lvscatter}, but for $b=\timeform{+0.30D}$ (b) Same as Figure \ref{lvXCO}, but for $b=\timeform{+0.30D}$}}  
\label{app+0.30}
\end{figure*}

 \begin{figure*}
\begin{center}    
\includegraphics[width=13.5cm]{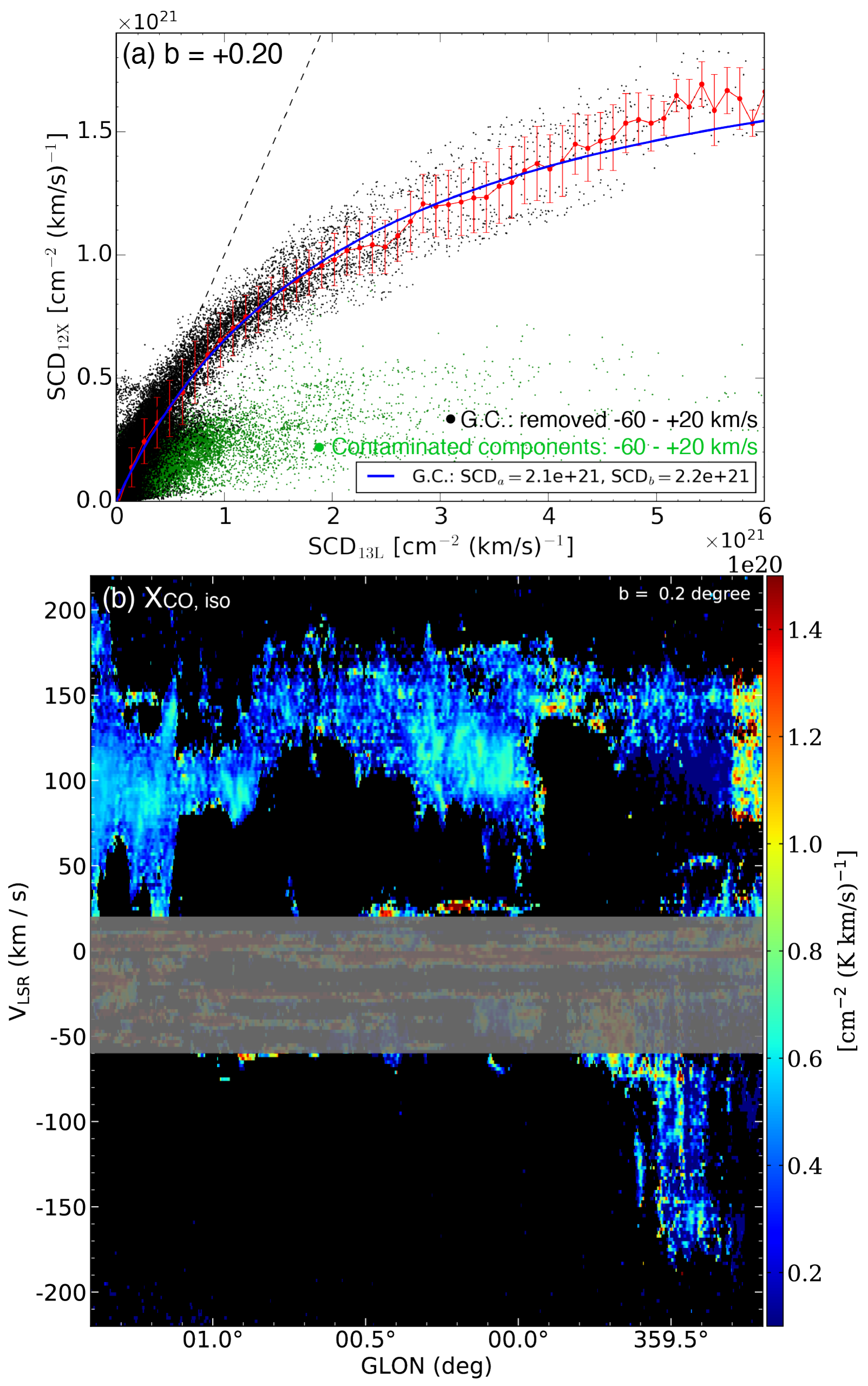}
\end{center}
\caption{{(a) Same as Figure \ref{lvscatter}, but for $b=\timeform{+0.20D}$ (b) Same as Figure \ref{lvXCO}, but for $b=\timeform{+0.20D}$}}  
\label{app+0.20}
\end{figure*}

 \begin{figure*}
\begin{center}    
\includegraphics[width=13.5cm]{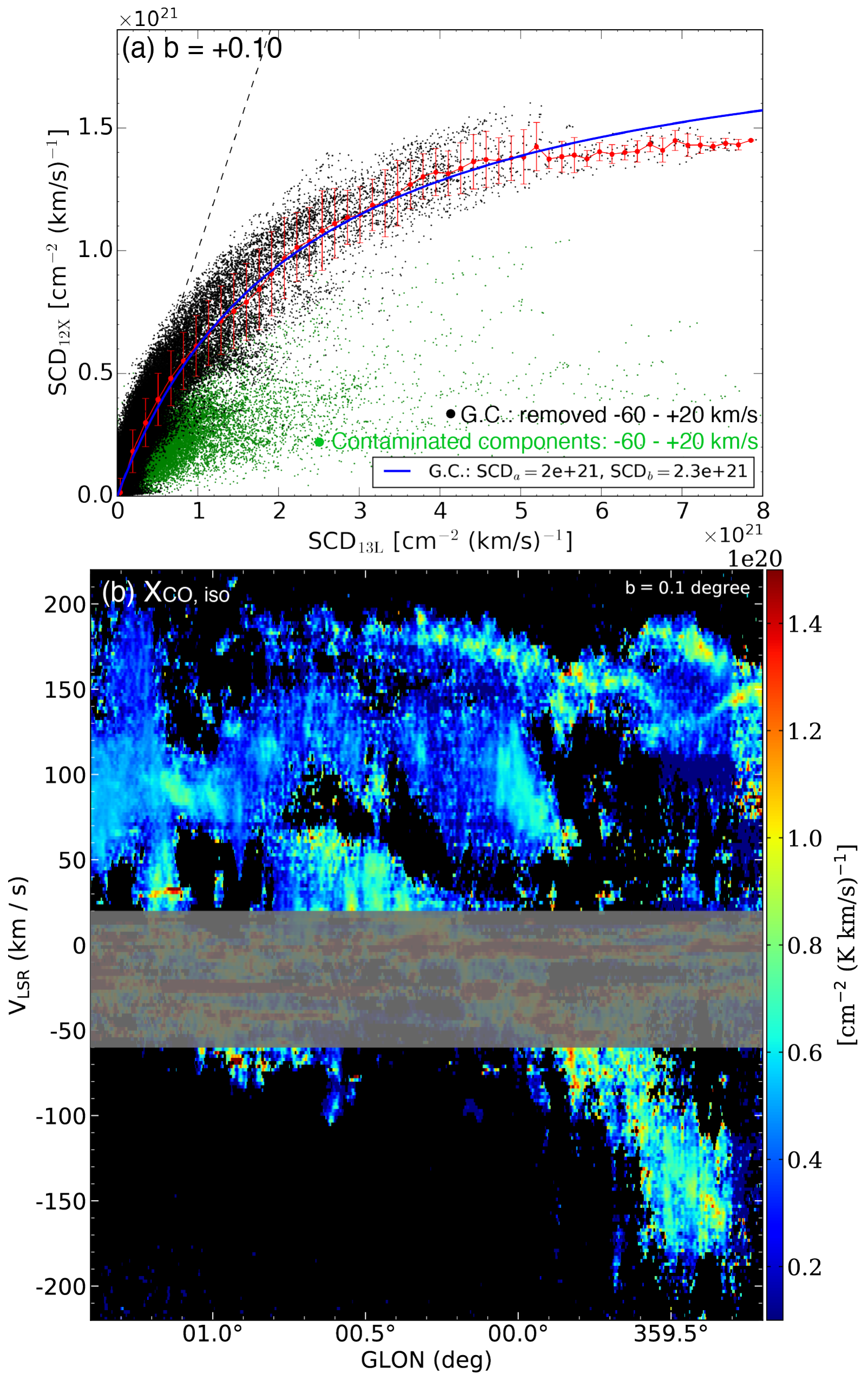}
\end{center}
\caption{{(a) Same as Figure \ref{lvscatter}, but for $b=\timeform{+0.10D}$ (b) Same as Figure \ref{lvXCO}, but for $b=\timeform{+0.10D}$}}  
\label{app+0.10}
\end{figure*}

 \begin{figure*}
\begin{center}    
\includegraphics[width=13.5cm]{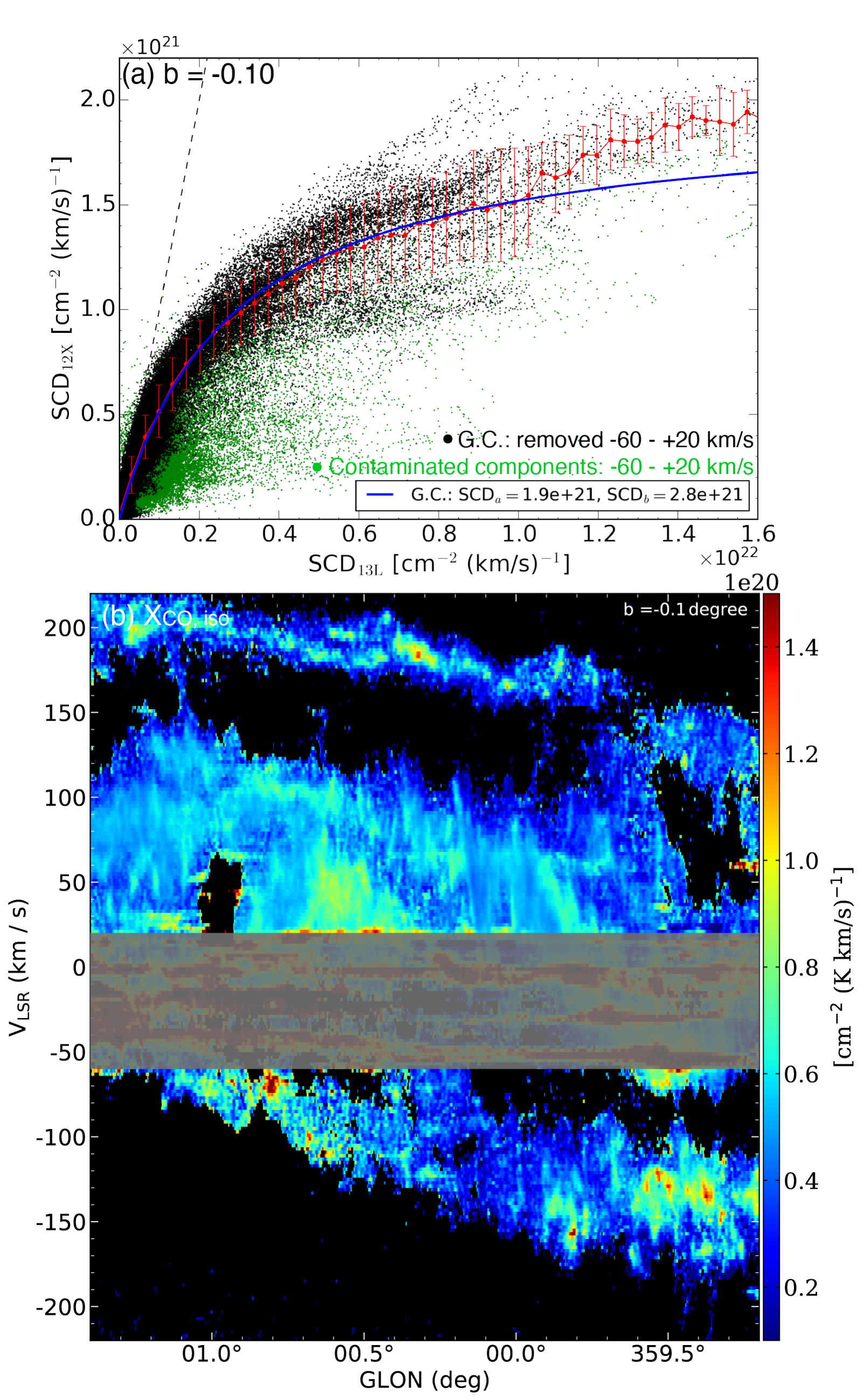}
\end{center}
\caption{{(a) Same as Figure \ref{lvscatter}, but for $b=\timeform{-0.10D}$ (b) Same as Figure \ref{lvXCO}, but for $b=\timeform{-0.10D}$}}  
\label{app-0.10}
\end{figure*}

 \begin{figure*}
\begin{center}    
\includegraphics[width=13.5cm]{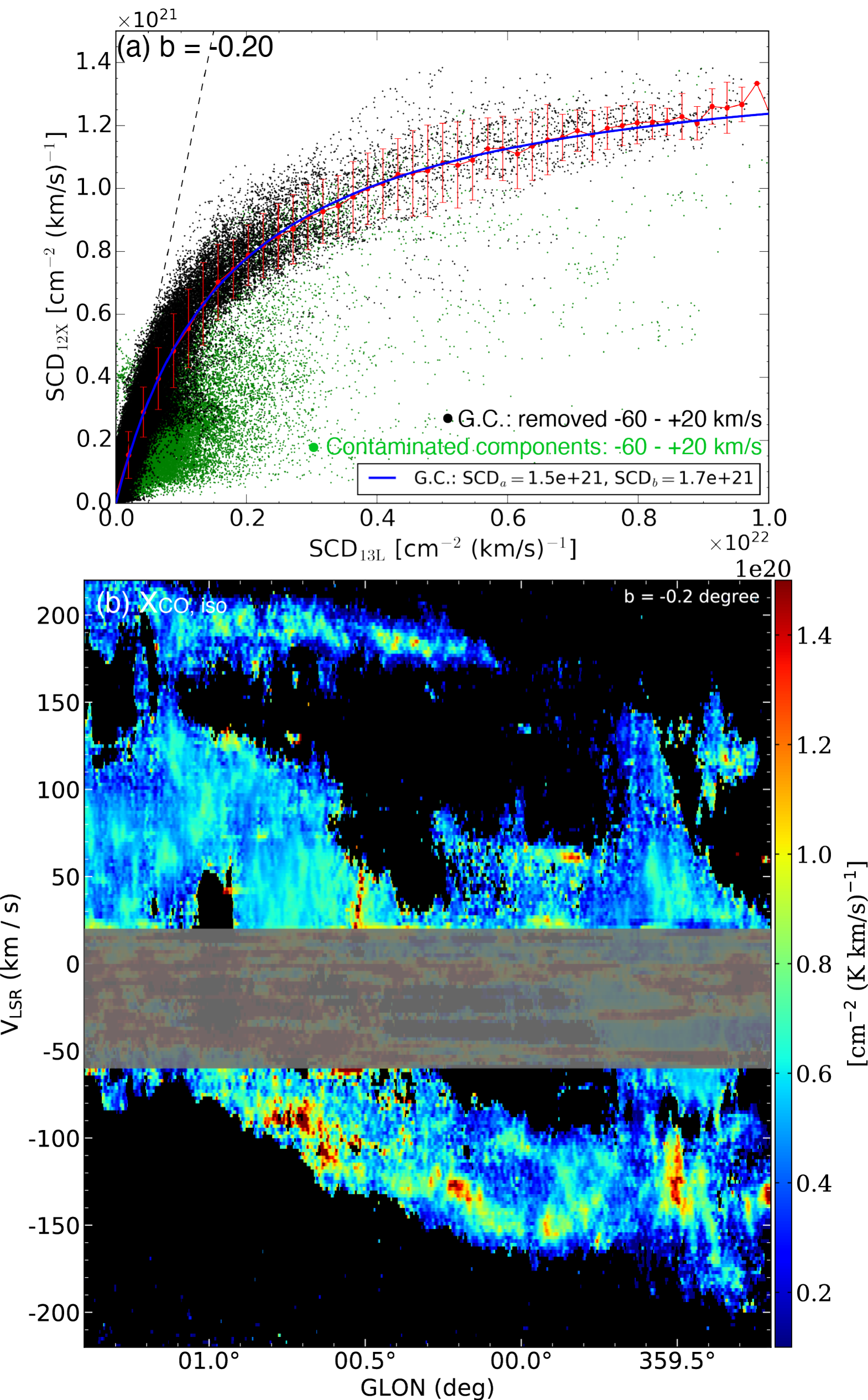}
\end{center}
\caption{{(a) Same as Figure \ref{lvscatter}, but for $b=\timeform{-0.20D}$ (b) Same as Figure \ref{lvXCO}, but for $b=\timeform{-0.20D}$}}  
\label{app-0.20}
\end{figure*}

 \begin{figure*}
\begin{center}    
\includegraphics[width=13.5cm]{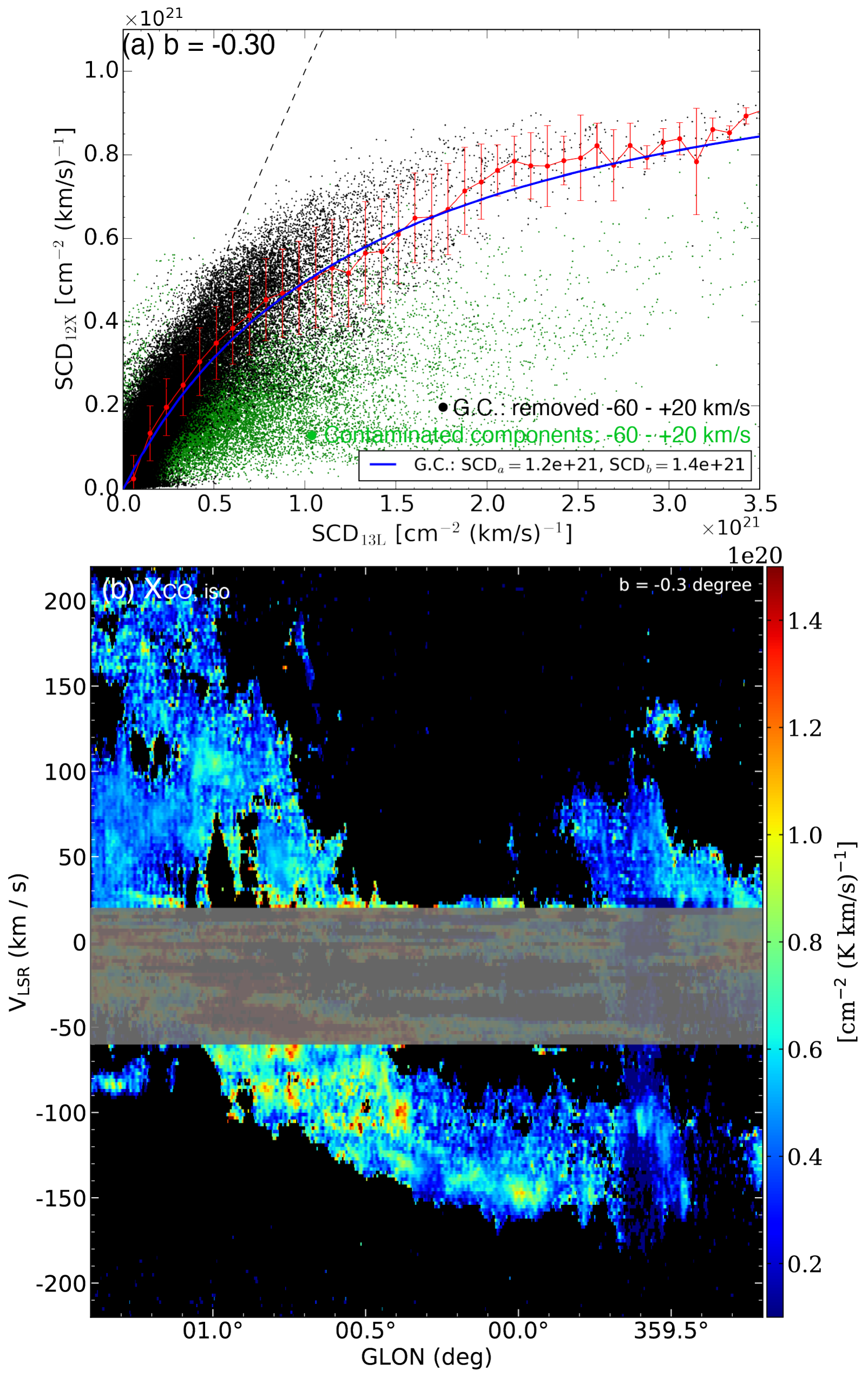}
\end{center}
\caption{{(a) Same as Figure \ref{lvscatter}, but for $b=\timeform{-0.30D}$ (b) Same as Figure \ref{lvXCO}, but for $b=\timeform{-0.30D}$}}  
\label{app-0.3}
\end{figure*}

\end{document}